\documentclass{svmult}

\usepackage{makeidx}         
\usepackage{graphicx}        
\usepackage{multicol}        
\usepackage[bottom]{footmisc}
\makeindex

\begin{document}

\title*{The International Trade Network}
\titlerunning{The International Trade Network}
\author{K. Bhattacharya$^1$, G. Mukherjee$^{1,2}$, and S. S. Manna$^1$}
\authorrunning{K. Bhattacharya, G. Mukherjee, and S. S. Manna}
\institute{Satyendra Nath Bose National Centre for Basic Sciences \\
           Block-JD, Sector-III, Salt Lake, Kolkata-700098, India \\
\and Bidhan Chandra College, Asansol 713304, Dt. Burdwan, West Bengal, India
\texttt{kunal@bose.res.in, gautamm@bose.res.in, manna@bose.res.in}}
\maketitle

\begin{abstract}
      Bilateral trade relationships in the international level between pairs of countries in the world 
   give rise to the notion of the International Trade Network (ITN). This network has attracted the 
   attention of network researchers as it serves as an excellent example of the weighted networks, the 
   link weight being defined as a measure of the volume of trade between two countries. In this paper 
   we analyzed the international trade data for 53 years and studied in detail the variations of different 
   network related quantities associated with the ITN. Our observation is that the ITN has also a scale 
   invariant structure like many other real-world networks.
\end{abstract}

      From long time back different countries in the world were dependent economically on many other countries
   in terms of bilateral trades. A country exports its surplus products to other countries and at the same
   time imports a number of commodities from other countries to meet its deficit. These bilateral trades
   among different countries in the world have given rise to the notion of the International Trade Network.
   In recent years studying the structure, function and dynamics of a large number of complex networks, 
   both in the
   real-world as well as through theoretical modeling have attracted intensive attention from researchers
   in multi-disciplinary fields \cite {Barabasi,Dorogov,Newman} in which ITN has taken an important
   position in its own right \cite {Serrano,Garlaschelli,Serrano2,Manna1,Serrano1}.
   The volume of trade between two countries
   may be considered as a measure of the strength of mutual economic dependence between them. In the language
   of graph theory this strength is known as the weight associated with the link \cite {Deo}. While simple graphical
   representation of a network already gives much informations about its structure, it has been observed
   recently that in real-world networks like the Internet and the world-wide airport networks the links have widely
   varying weights and their distribution as well as evolution yield much insight into the dynamical
   processes involved in these networks \cite {Barrat1,Barrat2,Manna,WL}.

      Recently few papers have been published on the analysis of the ITN.
   The fractional GDP of different countries have been looked upon as the `fitness' for the international trade.
   Links are then placed between a pair of nodes according to a probability distribution function of their fitnesses 
   \cite {Garlaschelli}. Also the trade imbalances between different pairs of countries, measuring the excess
   of export of one country to another over its import from the same country, are studied \cite {Serrano2,Serrano1}. 
   Using this method one can define the backbone of the ITN \cite {Serrano1}.

\begin{figure}[top]
\begin{center}
\includegraphics[width=12.0cm]{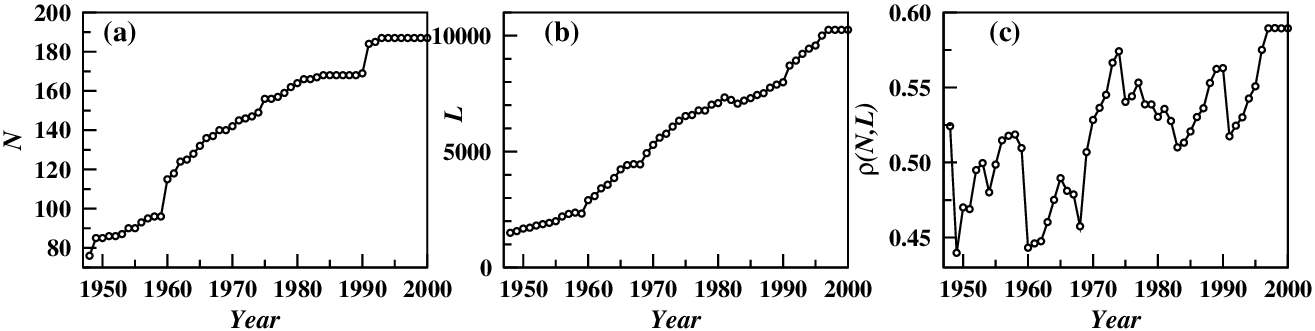}
\end{center}
\caption{
Variations of the (a) the total number of nodes $N$ (b) the total number of links $L$
and the (c) link density $\rho(N,L)$ of the annual ITN over a period of 53 years from
1948 to 2000. 
}
\end{figure}

      In the world there are different countries with different economic strengths. These
   countries are classified into three different categories. 
   According to the World Bank classification of different countries in July 2005 based on 
   gross national income (GNI) per capita as mentioned in the human development reports of 2003
   \cite {HDR} high income countries have GNI/capita at least $\$$9,386, middle income countries
   have GNI/capita in between $\$$9,386 and $\$$766 where as low income countries have
   GNI/capita less than $\$$766.

      In a recent paper \cite {Manna} we have studied the ITN as an example of the weighted networks. 
   Analysis of the ITN data over a period of 53 years from 1948 to 2000 available in \cite {Data} 
   have lead to the recognition of the following universal features: the link weight i.e., volume 
   of annual trade between two countries varies over a wide range and is characterized by a 
   log-normal probability distribution and this distribution remains robust over the entire period
   of 53 years within fluctuation. Secondly, the strength of a node, which is the total volume of 
   trade of a country in a year depends non-linearly with its Gross Domestic Product (GDP). In addition
   a number of crucial features observed from real-data analysis have been qualitatively reproduced
   in a non-conservative dynamic model of the international trade using the well known Gravity model 
   of the economics and social sciences as the starting point \cite {Tinbergen}.

      The annual trade between two countries $i$ and $j$ is described by four different quantities
   ${\rm exp}_{ij}$, ${\rm exp}_{ji}$, ${\rm imp}_{ij}$ and ${\rm imp}_{ji}$ measured in units of 
   million dollars in the data available in the website \cite {Data}. In general values of ${\rm exp}_{ij}$ 
   and ${\rm imp}_{ji}$ should be the same yet they have been quoted differently since exports 
   from $i$ to $j$ and $j'$s import from $i$ are reported as different flows in the IMF DOT data. 
   Although magnitudes of these quantities are approximately same in most cases they do differ 
   in many instances due to 
   different reporting procedures followed and different rates of duties applicable in different 
   countries etc. \cite {Gleditsh}. Therefore between two 
   countries $i$ and $j$ we denote the amount of export from $i$ to $j$ by $w^{exp}_{ij}$,
   the amount of import from $j$ to $i$ by $w^{imp}_{ij}$ and the total trade by $w_{ij}$ and 
   define them as:
\begin {equation}
w^{exp}_{ij}= \frac{1}{2}({\rm exp}_{ij}+{\rm imp}_{ji}),
w^{imp}_{ij}= \frac{1}{2}({\rm exp}_{ji}+{\rm imp}_{ij}),
w_{ij}= w^{exp}_{ij} + w^{imp}_{ij}.
\end {equation}

\begin{figure}[top]
\begin{center}
\includegraphics[width=12.0cm]{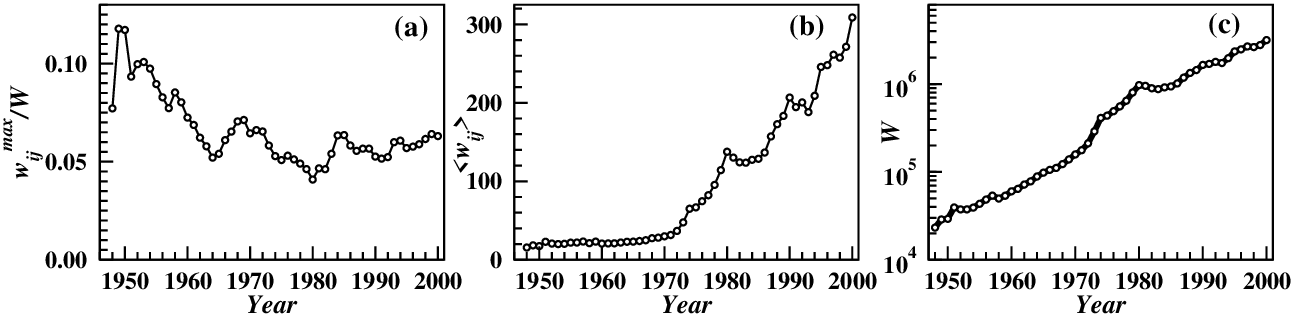}
\end{center}
\caption{
Plot of different quantities over the period from 1948 to 2000:
(a) the ratio of maximal trade $w^{max}_{ij}$ along a link and the total volume of trade $W$ of the ITN,
(b) average total trade (export + import) $\langle w_{ij} \rangle$ per link and
(c) total volume of annual world trade $W$.
}
\end{figure}

      Using these data the International Trade Network can be constructed every year. Naturally nodes 
   of the ITN represent different countries in the world. The export $w^{exp}_{ij}$ is the outword 
   flow from $i$ to $j$ and the import $w^{imp}_{ij}$ is the inward flow from $j$ to $i$. Therefore the 
   ITN is in general a directed graph with two opposite flows along a link, though it has been observed 
   that few links have only one flow. Obviously one can also ignore the direction and define an undirected 
   link between an arbitrary pair of nodes if there exists a non-zero volume of trade in any direction 
   between the corresponding countries. Both the number of nodes $N$ as well as the number of links $L$ 
   in the annual ITN varied from one year to the other. In fact they had grown almost systematically 
   over the years. For example, the number of nodes have increased from $N = 76$ in 1948 to 187 in 2000 
   (Fig. 1(a)), the number of links have increased from $L$ = 1494 in 1948 to 10252 in the year 2000 (Fig. 
   1(b)) where as the link density $\rho(N,L)=L/[(N(N-1))/2]$ fluctuated widely but with a with a slow 
   increasing trend around a mean value of 0.52 over this period (Fig. 1(c)).

     Looking at the available data few general observations can be made: Few high income \cite {HDR}
  countries make trades to many other countries in the world. These countries form the large degree hubs
  of the network. In the other limit, a large number of low income countries make economic transactions
  to few other countries. Moreover, a rich-club of few top rich countries actually trade among
  themselves a major fraction of the total volume of international trade.

   A huge variation of the volume of the bilateral trade is observed starting from a fraction of a million 
   dollar to million million dollars. There are a large number of links with very small weights and this 
   number gradually decreases to a few links with very large weights. The tail of the distribution consists 
   of links with very large weights corresponding to mutual trades among very few high income countries 
   \cite {HDR}. The variation of the ratio of $w_{max}$ and $W$ is shown in Fig. 2(a). The average weight 
   per link had also grown almost systematically from 15.54 $M\$$ in 1948 to 308.8 $M\$$ in 2000 (Fig.2(b)). 
   Again the total world trade $W$ had grown with years from $2.3 \times 10^{10}$ dollars in 1948
   to $3.2 \times 10^{12}$ dollars in 2000 (Fig.2(c)).

\begin{figure}[top]
\begin{center}
\includegraphics[width=10.0cm]{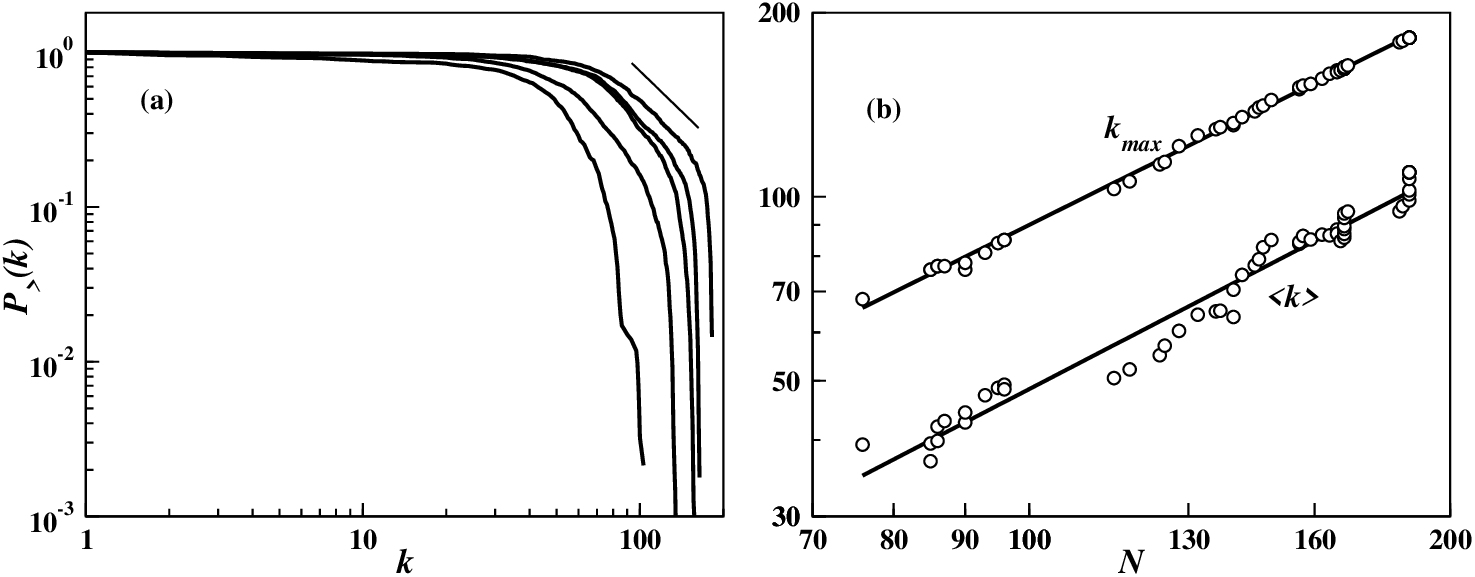}
\end{center}
\caption{
(a) Cumulative degree distributions $P_{>}(k)$ vs. $k$ averaged over the ten year periods
    during 1951-60, 1961-70, 1971-80, 1981-90 and 1991-2000 (from left to right). No power law 
    variation is observed for the 1951-60 plot. For the next decades however power laws
    over small regions are observed whose slopes gradually decrease to 1.74 for the 1991-2000 plot. 
(b) average nodal degree $\langle k \rangle$ and the largest degree $k_{max}$ with the size $N$
of the ITN.
}
\end{figure}

      The degree $k$ of a node is the number of other countries with which this country has trade 
   relationships. This can be further classified by the number of countries to which
   this country exports and is denoted by $k_{exp}$ where as $k_{imp}$ is the number
   of countries from which this country imports.
   In general $k_{exp} \ne k_{imp}$ but for some nodes they may be the same.
   The structure of the ITN is mainly reflected in its degree distribution which has been already
   studied in \cite {Garlaschelli} in which a power law for the cumulative distribution
   $P_>(k) \sim k^{1-\gamma}$ has been
   observed over a small range of $k$ values with $\gamma \approx 2.6$. We have studied the degree distributions,
   each averaged over ten successive ITNs, for example, 1951-60, 1961-70, 1971-80, 1981-90 and 1991-2000.
   The plots are given in Fig. 3(a). We see that indeed a small power law region appears for the period 
   1991-2000 with a value of $\gamma \approx 2.74$. Such a region is completely absent in the decade
   1951-1960. In the intermediate decades similar short power law regions are observed with larger
   values of $\gamma$. The average degree of a node $\langle k \rangle$ and the maximal degree
   of a node $k_{max}$ have also been studied for all the 53 years where the size $N$ of the ITN varied.
   We plot these quantities in Fig. 3(b) using double-log scale and observe the following power
   law growths as:
   $\langle k \rangle \sim N^{1.19}$ and $k_{max} \sim N^{1.14}$. Obviously these exponents have the
   upper bound equal to unity yet they are found out to be larger than one since both $\langle k \rangle/N$
   and $k_{max}/N$ ratios have grown slowly with time as as time progresses. This implies that as years
   have passed not only more countries have taken part in the ITN but in general individual countries 
   have established trade relationships with increasing number of other countries, a reflection of the
   economic global liberalisation.

\begin{figure}[top]
\begin{center}
\includegraphics[width=10.0cm]{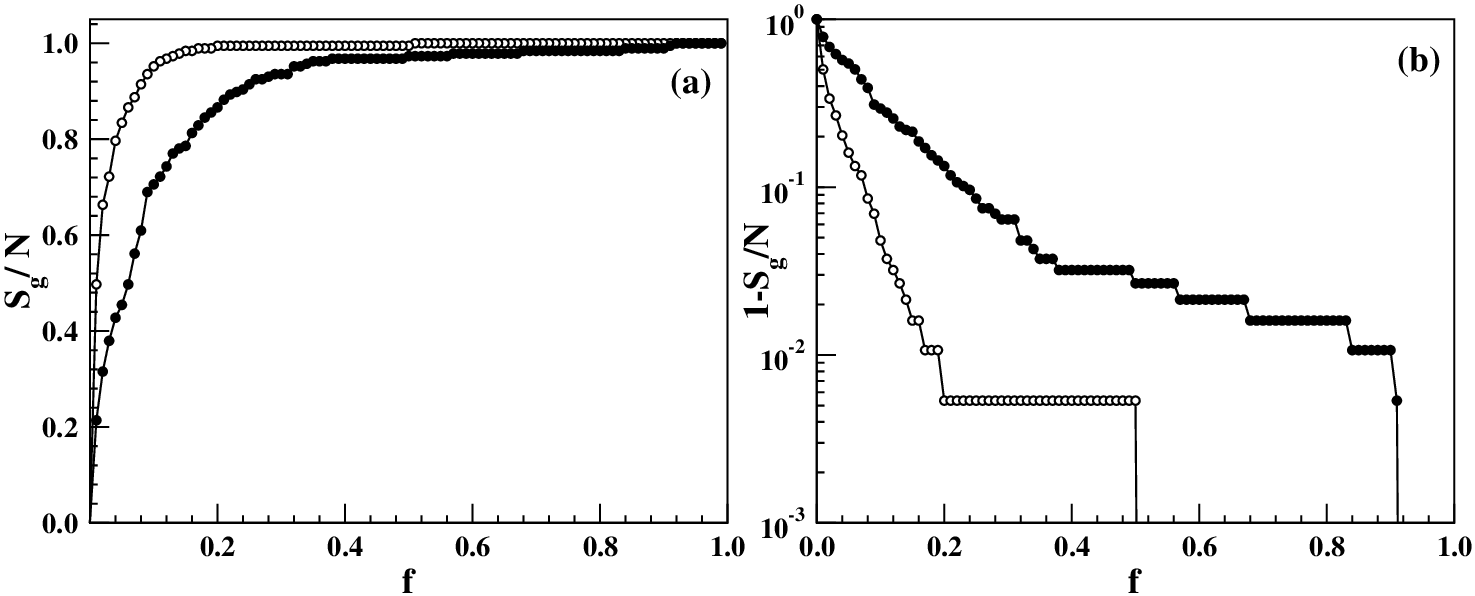}
\end{center}
\caption{
(a) Fractional size $S_g/N$ of the giant component of the ITN is plotted with the fraction $f$ of 
the links that are dropped in the descending (solid) sequence from the strongest and in the 
ascending (opaque) sequence from the weakest weight of the links. (b) The difference $1-S_g/N$ has 
been plotted on a semi-log scale which indicates an exponential approach to the fully connected network.
}
\end{figure}
      
      Within a year how the ITN grows to its fully connected configuration? A flavour of this mechanism can be
   obtained by the following mimicry. Consider a process which starts from $N$ nodes but with no links.
   Links are then inserted between pairs of nodes with a probability proportional to the weight of the
   link since a large weight link is more likely to be occupied than a small weight link.
   To do this first the link weights in the ITN have been ordered in an increasing sequence.
   Then the links are dropped in the descending order of the link weights starting from the maximum 
   weight $w_{max}$. We have also studied the reverse procedure when links are dropped in
   the increasing sequence of the link weights starting from the weakest link.
   In the Fig. 4(a) we show the growth of the fractional
   size of the giant component $S_g/N$ with the fraction $f$ of links dropped. The plot
   shows that the growth rate is slower in the first case and the giant component spans
   the whole ITN faster than when links are dropped in the ascending order of strengths.
   Moreover how the single connected component is attained has been quantitatively studied by plotting
   $1-S_g/N$ and $f$ on a semi-log scale in Fig. 4(b). The intermidiate straight portions
   in both plots indicate exponential growths of the size of the giant component.

      The annual volume of trade between a pair of countries is a measure of the strength
   of trade between them and is referred as the weight of the link connecting the 
   corresponding nodes. We have studied the distribution of the total trade $w_{ij}$
   along a link in detail, without distinguishing between the exports and the imports. 
   Therefore ${\rm Prob}(w)dw$ is
   the probability to find a randomly selected link whose weight lies between $w$ and $w+dw$.
   In general in a typical ITN, the link weights vary over a wide range. There are many many
   links with small weights whose number gradually decreases to a few links with large weights.
   In the first attempt we plot the distribution on a double logarithmic scale as shown
   in Fig. 5(a). Data for the six different years from 1950 to 2000 at the interval
   of ten years have been plotted with different colored symbols. Each plot has considerable
   noise which is more prominent at the tail of the distribution. Yet one can identify an intermediate
   region spanning little more than two decades of $w_{ij}$ where the individual plots
   look rather straight. This indicates the existence of a power law dependence of the 
   distribution: ${\rm Prob}(w) \sim w^{-\tau_w}$ in
   the intermediate regime. Therefore we measured the slopes of these plots in the
   intermediate region for every annual ITN for 53 years from 1948 to 2000. These values
   have fluctuations around their means and our final estimate for the exponent is: $\tau_w = 1.22 \pm 0.15$.

\begin{figure}[top]
\begin{center}
\includegraphics[width=12.0cm]{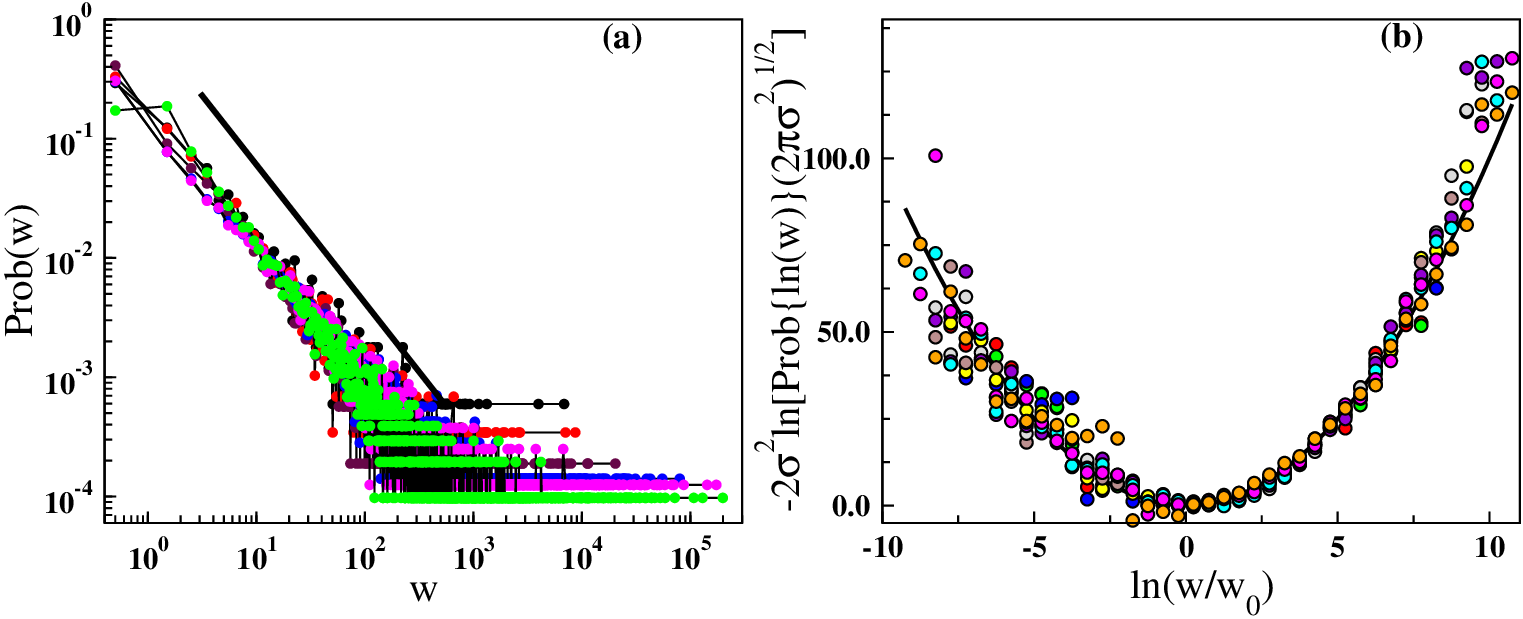}
\end{center}
\caption{(Color online) Trial of power law and log-normal fits:
(a) A double logarithmic plot of the probability distribution ${\rm Prob}(w)$ of the link weights
    for the six different years at the ten years interval from 1950 to 2000. The straight line
    shows average slope of the intermediate regime of all distributions giving an average estimate
    for the exponent $\tau_w = 1.22 \pm 0.15$.
(b) Scaled plot of the probability distribution of the link weights
$-2\sigma^2 \ln [{\rm Prob}\{\ln(w)\}\sqrt{2\pi\sigma^2}]$ as a function of $\ln(w/w_0)$.
The five year averaged data have been plotted for ten different periods from 1951 to 2000.
Points scatter around the scaled form of the log-normal distribution $y=x^2$ evenly
except at the ends.
}
\end{figure}
      
      We re-analyzed the same data by trying to fit a log-normal distribution as:
\begin {equation}
{\rm Prob}(w)=\frac{1}{\sqrt{2\pi\sigma^2}}{1\over w}\exp \left(-\frac{\ln^2(w/w_0)}{2\sigma^2}\right),
\end {equation}
where the characteristic constants constants of the distribution are defined as 
$w_0 = \exp(\langle \ln(w) \rangle)$ and
$\sigma = \{\langle (\ln(w))^2 \rangle - \langle \ln(w) \rangle ^2\}^{1/2}$. It is found
that different annual ITNs have different values for $w_0$ and $\sigma$. However
we observed that one can make a plot independent of these constants. Given the $w_{ij}$
values of an ITN one calculates first $w_0$ and $\sigma$. Then calculating the
${\rm Prob}\{ln(w)\}$ one plots
$-2\sigma^2 \ln [{\rm Prob}\{\ln(w)\}\sqrt{2\pi\sigma^2}]$ as a function of $\ln(w/w_0)$
which should be consistent with a simple parabola $y=x^2$ for all years (Note that
${\rm Prob}\{\ln(w)\}d\{\ln(w)\}={\rm Prob}(w)dw$ implies
${\rm Prob}\{\ln(w)\}=w{\rm Prob}(w)$). This analysis has been done for fifty years for the
period 1951-2000 but the data for every successive five years period have been averaged
to reduce noise and ten plots for the intervals 1951-55, 1956-60, ... , 1996-2000
have been plotted in Fig. 5(b) with different colored symbols. We observe that the data
points are evenly distributed around the $y=x^2$ parabola in most of the intermediate
region with slight deviations at the two extremes, i.e., at the lowest and highest values
of $\ln(w/w_0)$. We conclude that the probability distribution of link weights of the
annual ITNs is well approximated by the log-normal distribution and is a better 
candidate to represent the actual functional form of the ${\rm Prob}(w)$ than a power law. 
We mention here that
the trade imbalances have also been claimed to follow the log-normal distribution \cite{Serrano1}.

\begin{figure}[top]
\begin{center}
\includegraphics[width=12.0cm]{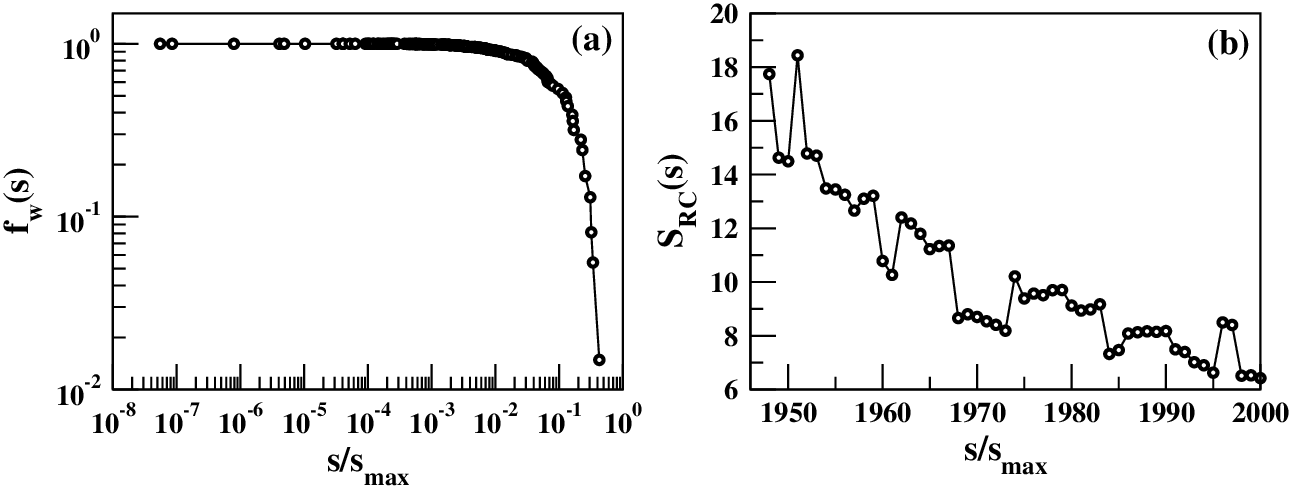}
\end{center}
\caption{
Variations of the 
(a) fraction of the total world trade the rich-club countries make among themselves
with the fractional strength of the weakest in the rich-club and
(b) percentage of the world countries that make 50$\%$ of the world's total trade
volume falls from $\approx19\%$ to $\approx6\%$ between 1948 to 2000.
}
\end{figure}
      
      In the world wide trade relations who is stronger and who is weaker? A measure
   of the capacity of trade is defined by the strength $s_i$ of a node which is the total sum
   of the weights $w_{ij}$ of the links meeting at a node $i$. Thus,
\begin {equation}
s_i=\Sigma_j w_{ij}.
\end {equation}
   Using the strength distribution one can estimate which countries actually control
   a major share of the international trade market. We define $f_w(s)$ as the ratio
   of the total volume of trade a subset of countries make among themselves to the
   total trade volume $W$ in the ITN. The subset is defined as those countries 
   whose strengths are at least $s$. For this analysis we first arrange the nodes
   in a sequence of increasing strengths and then delete the nodes in this
   sequence one by one. When a node is deleted all links meeting at this node are
   also deleted. Consequently the total volume of trade among the nodes in the subset
   also decreases. In the Fig. 6(a) we show how $f_w(s)$ decreases with $s/s_{max}$
   for the year 2000. Up to a large value of $s/s_{max} \approx 0.01$, $f_w(s)$
   effectively remains close to unity beyond which it decreases faster. It is 
   observed that only a few top rich countries indeed trade among themselves one half of the world's
   total trade volume, corresponding to $f_w(s)=1/2$. Evidently these countries are very rich 
   and are the few toppers in the list of strengths - which is said to have formed
   a `rich-club'(RC). Therefore we measure the fraction of countries in the RC
   and calculate how the percentage size of the rich-club varied with time. In Fig. 6(b)
   we plot the year-wise fractional size $S_{RC}$ of the rich club from 1948 to 2000
   and see that it has been decreased more or less systematically from $\approx 19\%$ to 
   $\approx 6\%$.
   This implies that though the world economy is progressing fast and more and more countries
   are taking part in the world trade market yet a major share of the total trade
   is being done only among a few countries within themselves.
    
\begin{figure}[top]
\begin{center}
\includegraphics[width=12.0cm]{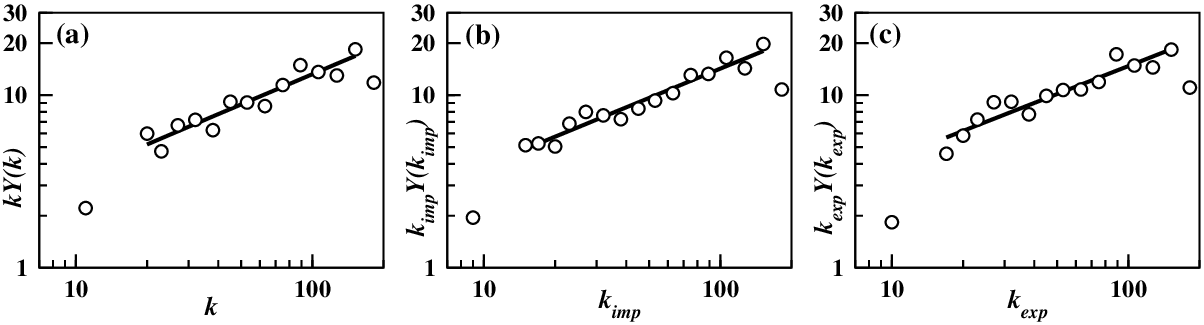}
\end{center}
\caption{
Binned plot of the nodal disparity measures with degree. 
Averaged ITN data for the period 1991 to 2000 have been used.
The best fit straight lines are 
(a) $kY(k) \sim k^{0.56}$ for the trade network,
(b) $k_{imp}Y(k_{imp}) \sim k_{imp}^{0.58}$ for the import network and
(c) $k_{exp}Y(k_{exp}) \sim k_{exp}^{0.54}$ for the export network.
}
\end{figure}

      A country makes different volumes of trade with other countries. Therefore the
   values of the weights associated with the links of a node, both for imports and 
   exports vary quite a lot. A numerical measure of this fluctuation is given by the
   `disparity' measure Y. For a node $i$ the disparity is measured by
\begin {equation}
Y_i=\sum^{k_i}_{j=1}\left[\frac{w_{ij}}{s_i}\right]^2.
\end {equation}
   The average disparity measure $Y(k)$ over all nodes of degree $k$ is calculated.
   If the weights associated with the $k$ links are of the same order then $Y(k) \sim 1/k$
   for large $k$ values where as if the weights of a few links strongly dominate over the 
   others then $Y(k)$ is of the order of unity. We have measured three disparity
   measures, namely $Y(k)$ for the link weights $w_{ij}$ as the total trade,
   $Y(k_{exp})$ for the link weights $w^{exp}_{ij}$ as the export from the node $i$ to node $j$ and
   $Y(k_{imp})$ for the link weights $w^{imp}_{ij}$ as the import from the node $j$ to node $i$.
   These quantities are plotted in Fig. 7 on log-log scales and we observe
   power law dependences as:
   $kY(k) \sim k^{0.56}$, $k_{imp}Y(k_{imp}) \sim k_{imp}^{0.58}$ and
   $k_{exp}Y(k_{exp}) \sim k_{exp}^{0.54}$. Similar variations are also observed in trade
   imbalances \cite {Serrano1}.

      To summarize, in this paper we have presented the analysis of the international
   countrywise trade data and studied the variations of different quantities
   associated with the International Trade Network, the trade data being available in \cite {Data}.
   While the ITN is inherently directed, where two opposite flows are associated with the majority of
   the links, we largely ignored the directedness and analyzed the network as an undirected graph.
   Our analysis shows that the link weight probability distribution of the undirected ITN fits
   better to a log-normal distribution as observed in \cite {Serrano,Serrano2}. We also show
   that the deviation in the size of the giant component of the ITN from the fully connected
   graph decays exponentially. The size of the rich-club whose internal trading amount is half of
   the total world trade amount decreases as time passes. Finally in the disparity measure of the
   ITN we distingushed between export and import at a node. It is observed that the three different
   disparity measures using link weights as the total trades, exports and imports grow in a
   similar fashion.

      We thank A. Chatterjee and B. K. Chakrabarti for their
   nice hospitality in the ECONOPHYS - KOLKATA III meeting.

\end {document}